\documentclass[12pt,a4paper]{article}

\usepackage[english]{babel}

\usepackage[utf8]{inputenc}

\usepackage[a4paper,top=2cm,bottom=2cm,left=2.5cm,right=2.5cm,marginparwidth=1.75cm]{geometry}

\usepackage[backend=bibtex, style=numeric-comp, sorting=none,defernumbers=true]{biblatex} 
\usepackage{amsmath}
\usepackage{amssymb}
\usepackage{amsfonts}
\usepackage{subcaption}
\usepackage{dsfont}
\usepackage{graphicx}
\usepackage[colorlinks=true, allcolors=blue]{hyperref}
\usepackage{hyperref}
\usepackage{orcidlink}
\usepackage{booktabs}
\usepackage{hhline}
\usepackage{multirow}
\usepackage[title]{appendix}
\usepackage{mathrsfs}
\usepackage{amsfonts}
\usepackage{booktabs} 
\usepackage{caption}  
\usepackage{threeparttable} 
\usepackage{algorithm}
\usepackage{algorithmicx}
\usepackage{algpseudocode}
\usepackage{longtable}
\usepackage{listings}
\usepackage{enumitem}
\usepackage{chngcntr}
\usepackage{booktabs}
\usepackage{lipsum}
\usepackage{subcaption}
\usepackage{authblk}
\usepackage[T1]{fontenc}    
\usepackage{csquotes}       
\usepackage{diagbox}

\newcommand{\be}{\begin{equation}}
\newcommand{\ee}{\end{equation}}

\usepackage{setspace}
\usepackage{titlesec}
\titleformat{\section} 
  {\normalfont\Large\bfseries}{\thesection.}{1em}{}

\usepackage{float}   
\usepackage{caption} 
\usepackage{hyperref} 

\hypersetup{
    colorlinks=true,
    linkcolor=black, 
    filecolor=black, 
    urlcolor=black   
}

\title{Defaultable bond liquidity spread estimation: an option-based approach}

\author[1,2]{Pietro Rossi\thanks{ \tt pietro.rossi3@unibo.it}}
\author[2]{Paolo Spezzati\thanks{\tt paolo.spezzati@prometeia.com}}
\author[2]{Riccardo Tedeschi\thanks{ \tt riccardo.tedeschi@prometeia.com}}

\affil[1]{\footnotesize  Department of Statistical Sciences "Paolo Fortunati", University of Bologna, Italy}
\affil[2]{\footnotesize  Pricing Unit and Financial Innovation, Prometeia S.p.a., Italy}

\date{\small This version: \today}  

\begin{document}
\maketitle

\begin{abstract}
This paper extends an option-theoretic approach to estimate liquidity spreads for corporate bonds. Inspired by Longstaff's equity market framework and subsequent work by Koziol and Sauerbier on risk-free zero-coupon bonds, the model views liquidity as a look-back option. The model accounts for the interplay of risk-free rate volatility and credit risk. A numerical analysis highlights the impact of these factors on the liquidity spread, particularly for bonds with different maturities and credit ratings. The methodology is applied to estimate the liquidity spread for unquoted bonds, with a specific case study on the Republic of Italy's debt, leveraging market data to calibrate model parameters and classify liquid versus illiquid emissions. This approach provides a robust tool for pricing illiquid bonds, emphasizing the importance of marketability in debt security valuation.
\end{abstract}

\bigskip


\newpage
{\small \tableofcontents}

\newpage
\section{Introduction}\label{sec:intro}

The concept of liquidity of financial assets and the risks associated with it is "nebulous and multi-faced" \cite{ilmanen2011expected}.
 In finance,  the term liquidity is commonly used in various fields: to characterize the overall monetary environment, to describe a condition of the balance sheet of an entity, the possibility of traders of funding their leveraged position and, in the financial market, the ease of trading. In this paper we are going to address the latter usage of the term liquidity and we will focus on a model to price such ease of converting an asset into money. In particular we will focus on a specific asset class, defaultable bond. 

Market liquidity, and its more evident effect, selling a bond, stems form the ability to trade and asset at low costs and with minimal price impact, allowing the holder to sell their holdings when needed. Usually market liquidity is characterized by its tightness (how wide bid-ask spreads are), the market depth (i.e the capability of  the market to absorb orders of a certain amount with little impact), and resilience (the reversion to equilibrium of prices after a shock). At this point we can already see how elusive the concept is. For example the market liquidity depth can be influenced by shortage of funding opportunity for market makers to meet the demands or it could be influenced by market sentiment towards particular investments due to specific market views. Analogous reasoning could be applied to tightness and resilience of the market. Therefore, it it paramount when dealing with models for pricing liquidity to explicitly highlight which aspects of liquidity are been considered.  
When dealing with fixed income markets, and in particular, with the bond market, the market liquidity is an important factor, given that this market is characterized by a lower number of transaction on corporate bond with respect to equity shares of the same entity. 

The impact of liquidity on bond prices and yields is well known in the literature and it has been gauged with many different liquidity measures. The different measures obviously gauge different aspects of liquidity and not always yield a unified picture. For example in \cite{bao2008excess} J. Bao and J. Pan measure liquidity as time covariance of the of price variation finding that this liquidity indicator predicts an higher illiquidity of the market than valuing liquidity using bond bid-ask spread. They conclude that bid ask spreads do not gauge depth and resilience of the market, that affect liquidity driven bond premia. Other examples of liquidity measure are the one of Y. Amihud \cite{AMIHUD200231}, defined as mean of yield changes in trades weighted by the size of the volume traded, and the liquidity measure of and Pastor and Stambaugh \cite{PastorStambaugh2003}, measuring temporary price changes  with order flow and their reversal speed. These measures, together with other liquidity indicators (on/off the run indicator, funding liquidity indicator), were exploited by H. Lin et Al. in \cite{lin2011liquidity} to estimate the liquidity impact on bond returns. 
They found that liquidity risk is in fact priced around 97 bps cross sectional on the bond market. Another example of market liquidity measure can be found  in \cite{Hu2010,Berenguer2013}, where it is defined as noise around the yield curve of Treasuries. Also liquidity indicators have been distilled from publicly available liquidity proxies (such as Treasury Bill vs. EuroDollar Futures - TED spread, LIBOR/OIS spread, VIX index, on-of spread of treasuries, swaptions volatility) performing a PCA analysis by R. Rebonato and H. Sherwin \cite{Rebonato2020}. The authors showed that the first two principal components can be traced to two different type of liquidity (funding liquidity and market liquidity, see \cite{Brunnermeier2008}) and therefore available proxies usually intertwines different liquidity concepts. Other indicators have been built of bid ask spread \cite{Konstantinovsky2016}. 
Despite the different measures and indicators listed do not gauge the market liquidity in coherent frameworks, measuring various aspects of liquidity, they all point to the relevance of liquidity assessment when valuing debt securities. This is more relevant when pricing unquoted bonds, which compose a huge volume of the whole market since debt emission are often traded OTC.  This is particularly important in the European market where it is relatively frequent to observe private placement to institutional investors, where a single emission is detained by few investors. 

Our contribution stems from the seminal work of F. A. Longstaff \cite{Longstaff1995}, followed by C. Koziol and P. Sauerbier \cite{Koziol2003} on non-marketability. The idea in these two papers is to view liquidity as the option or right to sell the security: an asset can be liquidated at any time or with different frequency whether is liquid or not. Therefore pricing liquidity is equivalent to price a derivative: in particular a "look-back option". Longstaff applied this framework to the equity market, while in   \cite{Koziol2003}  Koziol and Saurbier extended the framework to non defaultable zero coupon bonds. However as it has been established in the literature (for example \cite{bao2008excess}), the interplay of credit risk and liqudity risk has a considerable impact on bond yields, and the credit component is completely neglected by  C. Koziol and P. Sauerbier in  \cite{Koziol2003} .  
To overcome this limitation in \cite{Baviera2019}, Baviera et al. derived a closed formula for pricing illiquid corporate defaultable bonds focusing on the sheer liquidity, i.e. estimating the risk associated with bond price volatility and the time to liquidate a given bond position. In this paper we derive a liquidity discount neglecting the impact of liquidity on the bond prices, but getting an explicit and analytic estimation of the liquidity spread. Other attempts to take into account the interplay of liquidity and credit risk has been made in \cite{Chen2016,Li2021,Janosi2002}.

In this paper we will extend the numerical work of C. Koziol and P. Sauerbier in  \cite{Koziol2003} to include credit risk and coupon bond to their analysis. We will not derive an analytic formula but we are going to assess the interplay between credit risk and liquidity risks. We are going to study the sensitivity of our model to both changes in the rates volatility, and credit risks firstly for defaultable zero coupon bond and then we will extend the model for coupon bonds. We will then show a procedure to price illiquid unquoted bonds. In particular we will focus on emission of the Republic of Italy, which has a considerable amount of emission, most of them very liquid. The procedure we are going to set forth will comprise a classification of the relative liquidity of bonds and then we will calibrate the model for pricing unquoted bonds.

We observe that the idea of pricing and illiquid product is somewhat questionable, what we aim is estimating a range of prices that take into account liquidity. This is a limitation of all the model that deal with liquidity, given that illiquid markets entails incomplete information.

The paper is organized as follows: in Section \ref{sec2} we are going to describe the model and the theoretical framework. In Section \ref{sec3} we will discuss the features and sensitivity of the model to the parameters. In Section \ref{sec4} we will study the calibration of model parameters to price unquoted emission of the Republic of Italy. The procedure outlined could be also used for pricing corporate bonds.

\section{The model}\label{sec2}

Following the seminal paper on price of marketability of Longstaff \cite{Longstaff1995}, and the subsequent work extending his ideas of Koziol and Sauerbier \cite{Koziol2003} to default-free zero coupon bond, we are going to illustrate a further extension that includes defaultable coupon bonds. Our work differs from other reinterpretation of the principles of Longstaff that other authors have exploited to compute sheer liquidity as in \cite{Baviera2019}, where the back-reaction impact of liquidity is neglected. 

Before discussing the model and the principles underneath it, let us fix the notation used. 

\subsection{Notation}\label{subsec21}

First of all, we will denote the stochastic discount factor and its expected value with respect to the usual filtration $\mathcal{F}_t$ generated by the stochastic processes considered as
\begin{equation}
D(t,t_1)=e^{-\int_{t}^{T}r_s ds}, \qquad P(t,T) = \mathbb{E}[D(t,T)|\mathcal{F}_t]=\mathbb{E}_{t}[D(t,T)].
\end{equation}
 The uncertainty of the model is characterized by the complete filtered probability space $(\Omega,\mathcal{F}, (\mathcal{F}_t)_{t \in [0,T]}, \mathbb{Q})$\footnote{See Protter \cite{Protter2003} for the definitions of the various terms}. Here we assume that the probability measure is what is called the martingale  measure, that is assumed to exist\footnote{with very few hypothesis it can be shown to exists, see \cite{Protter2003, Musiela1998}}. 
In this notation the value\footnote{When $t>0$ then $B(t,T)$ is a stochastic variable.} of a default-free fixed-coupon bond with maturity $T$ can be written as
\begin{equation}
B(t, T) = \sum_{t_i>t}^T c_iP(t,t_i),
\end{equation}

where$ \{t_i\}_{i=1,\dots,N} \text{ are the payment dates}, c_i$ are the coupons paid at each payment dates (for $t_N=T$ the coupon include also the notional capital).
 
We are going to model default using a reduced-form model. Firstly we will assume that the credit risk intensity process and the risk-free interest rate process will be uncorrelated  as in \cite{Duffie1998DefaultableTS,Hilscher2018} (for correlated processes see  \cite{Duffie1999} or \cite{Schonbucher1997} where they consider correlated models of HJM family). Let $\lambda: [0,T] \times \mathbb{R}^n\rightarrow [0,\infty)$ be the intensity process of a Cox process $N_t \in\{0,1,2,\dots\}$ with $N_0=0$. Here the filtration of the probability space is $\mathcal{G}_t=\mathcal{F}_t\vee\mathcal{H}_t$ where the $\mathcal{H}_t$ is the filtration generated by the point process and  $\mathcal{F}_t$ is the filtration generated by the intensity process and the risk-free rate process. Now we define the stopping time representing the firm's default as :
\begin{equation}
\tau = inf\{t>0: N_t=1\},
\end{equation}
therefore the intensity $\lambda_t$ can be interpreted as the intensity of default or the default probability in the time interval $[t, t+dt]$ conditional to survival up to $t$.
Given this we can write ZC defaultable bond  as 
\begin{equation}
\bar{P}(t,t_1)=\mathds{1}_{\{\tau>t\}} \mathbb{E}_{t}[D(t,t_1)\mathds{1}_{\{\tau>t_1\}} + RR_{\tau}  \mathds{1}_{\{\tau<t_1\}}D(t,\tau)]
\end{equation}
We simplify further the model considering a constant recovery rate $RR_t=RR$.
Under the assumption that the default-free spot rate $r_t$ and the default time $\tau$  are independent under $\mathbb{Q}$ we can rewrite the ZC bond as\footnote{Here the formulas assumes that theZC bond issuer has not defaulted before time $t$}
\begin{align}
&\bar{P}(t,t_1)=\mathds{1}_{\{\tau>t\}} \bigl[z(t,t_1) +RR \cdot d(t,\tau)\bigr] \\
&z(t,t_1) =\mathbb{E}^{\mathbb{Q}}[\mathds{1}_{\{\tau>t_1\}}D(t,t_1)|\mathcal{F}_{t}] = P(t,t_1) \cdot \mathbb{E}^{\mathbb{Q}}_{t}[\mathds{1}_{\{\tau>t_1\}}]\\
&d(t,\tau) = \mathbb{E}^{\mathbb{Q}}[\mathds{1}_{\{\tau\le t_1\}}D(t,\tau)|\mathcal{F}_{t}] 
\end{align}
where $z(t,t_1)$ is a the price of zero coupon bond with zero recovery cashed in only case of survival and $d(t,t_1)$ is the price of a digital default bond (i.e. a zero coupon bond cashed in only in case of default).  Now we can define the survival probability as 
\begin{equation}
S(t, T) = \mathds{1}_{\{\tau>t\}} \mathbb{E}^{\mathbb{Q}}_t[\mathds{1}_{\{\tau<T\}}] = \mathds{1}_{\{\tau>t\}} \mathbb{E}^{\mathbb{Q}}_t[e^{- \int_t^T\lambda_s ds}].
\end{equation}
Appropriately rescaling the intensity process $\lambda_t\rightarrow s_t=\lambda_t LGD$, where  $LGD = (1-RR)$ is the Loss Given Default (complement to unit of the Recovery Rate) expressed as a percentage of the market value just before default, one can write the risky discount factor $\bar{P}(t,T) =\mathds{1}_{\{\tau>t\}} \mathbb{E}^{\mathbb{Q}}_t[e^{-\int_t^T (r_s+s_s) ds}]$\footnote{see \cite{Lando1998} or \cite{Duffie1999} for derivation of this formula.}. 
Viewing a fixed coupon bond $V^B$ with  maturity $T$ and unit notional as a portfolio of zero coupon bond we get its value at time $t$ as:
\begin{equation}
V^B(t, T) = \sum_{i=1}^{N}c_i \bar{P}(t,t_i)
\end{equation}
 where $T$ is the maturity and $\{(c_i,t_i)\}_{i=1,\dots,N}$ are the coupons and the payment dates time respectively. 

Corporate bond are illiquid compared to treasury or other equities. It has been shown that the returns of corporate bond comprise an illiquidity discount. In the paper we will assume that this liquidity discount $\gamma$ is constant and that it modifies the value of the corporate bond in the following way:
\begin{equation}
V^B(t,T)= \sum_{i=1}^{N}c_i e^{-\gamma(t_i-t)}\bar{P}(t,t_i). 
\end{equation} 
The purpose of this paper is to estimate the value of $\gamma$ given different liquidity features of the bond. For simplicity we define $\tilde{P}(t,t_1)= e^{-\gamma(t_1-t)}\bar{P}(t,t_1)$ as the price of a defaultable (risky) zero coupon bond liquidity adjusted.  

\subsection{The option-theoretic approach to illiquidity}\label{subsec22}

In this section we present the main idea, similar to \cite{Longstaff1995} and developed by \cite{Koziol2003}, for the option-theoretical approach to calculate liquidity spread  for rarely traded bond or private debt.  This approach links look-back option\footnote{Strictly speaking, we will deal with a "strike zero" option, which is not a fully-fledge option. Nonetheless, we will refer to this contingent claim as a look-back option, following the name that appears in the literature.} value with the benefits from liquidity in a similar way Longstaff valued the marketability of stocks. The benefits of liquidity are gauged looking at the potential trading gains of an investor with perfect market ability. Here this ideal investor, which has maximal information, is able to sell the security and reinvest the proceeds at the risk-free rate in order to maximize the value of its portfolio at maturity. This ideal investor is a "know-all" investor, that is able to harness the difference between the price of the security at sale time and its real value, given the realized evolution of the market variables. This ability translates in defining (and finding) the maximum of the disposal strategy only at maturity of the option (that will coincides with the bond maturity), this identifies the look-back feature of the option. The link with the benefits from liquidity is then established computing the value of this option in two different markets: a liquid market where the sale of the security can be performed at any time, and an illiquid case in which the sale can happen only on a discrete set of dates. The different constraint on the sales dates, or observation point allowed in the calculation of the payoff of the option, gauge the effects of illiquidity: the fewer observation instances, the lower the option value.  The liquidity discount is  inferred from the ratio of the option value in these two different markets. 

Let us illustrate the risk factors involved and the option payoff. Considering defaultable zero coupon (ZC) bond, the risk factors involved in their analysis are the risk free interest rates and the credit spreads. The risk-free rates are modelled using a short rate $r_t$ stochastic model (without loss of generality we will consider short rates model, but the same arguments can be applied to other frameworks). The credit spreads are modelled by a Cox process $\lambda_t$. 

We consider therefore two market in which two bonds with the same characteristics can be traded with different frequencies. The set of possible traded date will be denoted by $\mathcal{T}$. In the liquid market, this set is equal to $\mathcal{T}=\mathcal{T}_i=[0,T]$\footnote{The continuum limit will be approximated using fifteen minutes trading frequency}. Instead, when we are dealing with the illiquid case, the set $\mathcal{T}=\mathcal{T}_{il}=\{t_i\}_{i=1,\dots, N}$ comprises only a finite set of dates. The information on liquidity is therefore characterized by the properties of this set. In general the distribution of these dates, provided that $0\le t_1<\dots<t_N =T$, can be arbitrary, and it has been shown in \cite{Koziol2003} that given the same number of dates, different distributions in the timeline of those dates lead to different liquidity spreads. In the following, for simplicity we will assume that the dates are distributed uniformly, which allows us to define the quantity $\Delta t =t_{i+1}-t_i\,\,\, \forall i$. We note that the fact that these bonds trade in different markets ensures the absence of arbitrage in these markets: \emph{this prevents hedging the liquid option with illiquid bonds and vice versa, allowing for arbitrage opportunity}.

In both these cases, liquid and illiquid, we consider the look-back option with payoff:
\begin{equation}
O(T,T) = \max\limits_{t\in \mathcal{T}}\biggl( \tilde{P}(t,T)e^{\int_t^Tr_s ds}\mathds{1}_{\{t<\tau\}}\biggr),
\end{equation}
where $\tilde{P}(t,T)=e^{-\gamma (T-t)}\bar{P}(t,T)$ has either $\gamma=0$ or $\gamma\neq 0$ whether we are considering the liquid case $\mathcal{T}=[0,T]$ or the illiquid case $\mathcal{T}=\{t_i\}_{i=1,\dots, N}$. Note that the Cox process affects the payoff of the look-back option in two ways. First, $\tilde{P}(t,T)$ depends on the intensity process $\lambda_t$ and models the changes of the not defaulted risky bond. Whereas the point process modeling defaults affects the payoff of the option, truncating the series of observation on which the maximum is estimated, restricting the possible observation to those where a default event did not happen.

As already noticed the sale time of the ZC bond $t$ maximizing the option strategy  is determined at maturity of the option. This precisely resembles the characteristic of our know-all investor. 

Using risk-neutral valuation, the value of the option at time zero is given by:
\begin{align*}
O(0,T)&=\mathbb{E}_0[D(0,T)O(T,T)]\\
&=\mathbb{E}_0\biggl[e^{-\int_0^Tr_sds}\max\limits_{t\in \mathcal{T}}\biggl( \tilde{P}(t,T)e^{\int_t^Tr_s ds}\mathds{1}_{\{t<\tau\}}\biggr)\biggr] \\
&= \mathbb{E}_0\biggl[\max\limits_{t\in \mathcal{T}}\biggl( \tilde{P}(t,T)e^{-\int_0^tr_s ds}\mathds{1}_{\{t<\tau\}}\biggr)\biggr]
\end{align*} 
Having determined the value of the look-back option we can define the liquidity discount adjustment factor as the value
\be
\label{eq:liqZC}
e^{-\gamma (T-t)} = \frac{O(0,T)\,\, valued\,\, on\,\, \mathcal{T}=\{t_i\}_{i=1,\dots,N}}{O(0,T)\,\, valued\,\, on\,\, \mathcal{T}=[0,T]}= \frac{\tilde{P}(0,T)}{\bar{P}(0,T)},
\ee

where $\gamma$  is the theoretical liquidity spread.

Notice that the first equality is the definition, while the second equality is in turn the underlying assumption of our approach. One can argue, as both \cite{Koziol2003, Longstaff1995} do, that the second equality is an upper bound and that the real liquidity discount is in fact a fraction $\gamma^{adj}=c\cdot\gamma$ of the estimated discount. This observation stems from the fact that computing the value of the look-back option entails considering an ideal "know all" investor, and that in the usual market condition the timing ability of market participants is sub-optimal (usually investors are not omniscient). Nevertheless we are going to determine and study the calculation of $\gamma$, without bothering too much to study the feature of the constant $c$ which should be another parameter that is market dependent. A possible calibration method has already been discussed in \cite{Koziol2003}.
 
Furthermore we are going to generalize the analysis of \cite{Koziol2003} extending the model for coupon bearing bonds. In this case one has to consider the maximization of the strategy that, not only consider the value of the bond
\be \label{eqRiskyBondValue}
 V^B(t,T) = \sum_{\{i | t_i>t\}}^Nc_i\tilde{P}(t,t_i)
\ee 
at a certain time but also the coupon already paid between the time zero and $t$. One has therefore to rewrite the payoff of the option at maturity $T$ as 
\begin{align*}
O(T,T) &=\max\limits_{t \in \mathcal{T}}\biggl\{\biggl(\sum_{i| t_i< t} c_i e^{\int_{t_i}^tr_s ds} \mathds{1}_{\{t_i<\tau\}}+ V^B(t,T)  \mathds{1}_{\{t<\tau\}} +RR \cdot V^B(\tau,T) e^{\int_{\tau}^tr_s ds} \mathds{1}_{\{t>\tau\}}\biggr)e^{\int_t^Tr_s ds}\biggr\}.
\end{align*}
Then substitutibg $V^B$ from equation \eqref{eqRiskyBondValue} one gets:
\begin{align*}
O(T,T) &=\max\limits_{t \in \mathcal{T}}\biggl\{\sum_{i=1}^N c_i \biggl[\tilde{P}(t,t_i)(\mathds{1}_{\{t_i>t, \tau>t\}}+\eta(\tau,t) \mathds{1}_{\{t_i>t, \tau<t\}})+e^{\int_{t_i}^tr_s ds}\mathds{1}_{\{t_i<t,\tau>t_i\}}\biggr] e^{\int_t^Tr_s ds}\biggr\}\\
 \text{where  }\eta(\tau, t)&=RR \cdot \tilde{P}(\tau,t) e^{\int_{\tau}^tr_sds}
\end{align*}
Risk-neutral valuation yield the present value of the option:
\be\label{eqlook-backOpt}
O(0,T) =\mathbb{E}_0\biggl[ \max\limits_{t \in \mathcal{T}}\biggl\{\sum_{i| t_i< t}  c_i e^{-\int_{0}^{t_i}r_s ds} \mathds{1}_{\{t_i<\tau\}}+ V^B(t,T)e^{-\int_0^tr_s ds}\mathds{1}_{\{t<\tau\}}+RR\cdot V^B(\tau,T) e^{-\int_{0}^{\tau}r_s ds} \mathds{1}_{\{t>\tau\}}\biggr\}\biggr]
\ee
Equation \eqref{eqlook-backOpt} represents the value of the look-back option as the sum of the (present) value of the coupons cashed up to selling time $t$, plus the value of the bond at time $t$ in case of survival, plus the recovery on the bond in case of default before time $t$, where $t$ is the time that that maximizes the strategy value. 

As in the case of zero coupon bond we assume that the impact of the different liquidity is captured by the difference of the option value considering the two sets of trading dates $\mathcal{T}_l$ and $\mathcal{T}_{il}$, i.e.
\be
\label{eq:liqbond}
\frac{O(0,T)\,\, valued\,\, on\,\, \mathcal{T}_{il}=\{t_i\}_{i=1,\dots,N}}{O(0,T)\,\, valued\,\, on\,\, \mathcal{T}_{l}=[0,T]} = \frac{V^B(0,T)}{V^B_{\gamma=0}(0,T)}
\ee
where we have denoted $V^B_{\gamma=0}(0,T)$ the value of the bond with $\gamma=0$. Note that both the equations \eqref{eq:liqZC} and \eqref{eq:liqbond} define and implicit equation for the parameter $\gamma$. In the next section we are going to solve those equations numerically. 

At this point we note that, in general, there is not an analytic formula  to compute the value of the look-back option. In special cases an analytic formula can be found especially for the liquid case. An example is the one considered by \cite{Koziol2003}, where they consider non-defaultable ZC  coupon bond and risk-free rates are modelled via a linear short-rate model (Vasicek model). In that case the probability distribution of the short rate is normal and the exponential of the short rate can be computed in closed form. This is still the case if one considers an Hull and White model, but it is not true any more for others commonly used short rate models. Therefore we are not going to focus on solving analytically the equations but we are going to perform a numerical analysis. 

\subsection{The Numerical Analysis}\label{secnumerical}

The numerical analysis uses the model of G2++ for simulating risk-free rates and a CIR model for simulating the credit spreads \cite{Brigo2006}. In  the G2++ model the short rate dynamics is defined by the following Stochastic differential equations:
\begin{align*}
r_t&=x_t+y_t+\phi_t, \quad r(0)=r_0, \quad x(0)=y(0)=0 \\
dx_t &=-ax_tdt+\sigma dW_{1t}\\
dy_t &=-by_tdt+\eta dW_{2t}, \qquad dW_{1t}dW_{2t}=\rho dt, 
\end{align*}
where $r_0, a,b,\sigma,\eta$ are positive constants, $\rho\in [-1,1]$ and the function $\phi$ is deterministic and $W_{1t}, W_{2t}$ are two Brownian motion.  
The risk-free short rates  model ensures perfect fitting with the term structure of risk-free rates  and still maintains simplicity. The model is calibrated using the swaption quotes . 
The credit spread $(s_t)$ dinamic is described by the following SDE
\be
ds_t = \kappa(\theta-s_t)dt+\sigma \sqrt{s_t}dW_t 
\ee
Credit spread model parameters are calibrated on the term structure of credit spread at reference date, while retaining the estimation of the volatility parameter $\sigma$, from a maximum likelihood estimation using the historical time series of credit spread w.r.t the euribor 6M. 

Then the value of the options are computed using a Monte Carlo Simulation. The liquid case considers look-back options that allow probing hourly the value of the strategy. Whereas in the illiquid case the allowed probing days are the set of dates $\mathcal{T}_{il}$, for different $\mathcal{T}_{il}$. As we will see, one of the analysis dimension is the probing frequency $\Delta t $: we are going to consider $\Delta t$ equals to one or two day, five days, ten days, twenty days, sixty days and hundred-twenty days.  Since the eq.\eqref{eq:liqbond} is an implicit equation in the unknown $\gamma$ we use the Newton algorithm to find the root of the equation. This method is very efficient given that the look-back option value is monotonic in the value of $\gamma$. 

Given the considerable impact of default event in computing the option value, one should consider a huge number of Monte Carlo trajectory for estimating $\gamma$, especially for investment grade bonds.  To avoid performing optimization with too many trajectories, we will set the number of trajectory $N=10000$ and then repeatedly estimate $\gamma$ for $m$ times, obtaining a sample of its values. The parameter $\gamma$ is then estimated as the mean of the sample.

\section{Analysis and results of defaultable zero coupon}\label{sec3} 

Let us turn to the analysis of the liquidity spread estimation for defaultable ZC bond. The aim is to study the impact on the liquidity bond estimation of the credit risk, trying to disentangle the impact of the default events from the impact of credit spread model. 
To analyse the various impact we will switch on and off the stochastic nature of these risk factors, weighting for each case their impact on the liquidity spread estimation. The model parameters are summarized in the appendix A in Table \ref{table:CIRparameter} for the credit spread model (we consider the BB rating parameters), while Table \ref{table:g2ppmodel} shows the parameter of the risk-free rates model. To perform such analysis we will fix the probing frequency of the illiquid option payoff to 5 days. The analysis will also uncover the impact of considering different maturities for the ZC bond:  for this reason we will consider ZC bond with maturity between 3 months and twenty-seven years. 
Schematically the analysis will consider the following scenarios
\begin{itemize}
\item[1.]\textbf{Case 1 - Risk-free rates}: we consider the following option payoff
\be
O(T,T)=\max\limits_{t \in \mathcal{T}}\biggl\{\biggl(\sum_{i| t_i< t} c_i e^{\int_{t_i}^tr_s ds} + V^B(t,T) \biggr)e^{\int_t^Tr_s ds}\biggr\}
\ee
 where $V^B(t,T)=\sum_{i|t_i>t}^Nc_i\tilde{P}(t,t_i)$. $V^B$ can be seen as a (non-defaultable) zero coupon bond whose value depends on deterministic credit spread, a constant liquidity spread, and risk-free interest rates modelled using a G2++ model. In this case the stochastic behaviour is driven solely by the risk-free rates part. Note that we omit the possibility of default event. In other words it is as if we consider the model of \cite{Koziol2003}, with a different assets as a risk-free bond. 
\item[2.] \textbf{Case 2 - Risk-free rates and credit spread}:  this case considers the same payoff as above, but it allows for a stochastic credit spread model. In other words the stochastic variables $\tilde{P}(t,t_i)$ determining the bond values depend on the stochastic processes of the risk-free interest rate and the credit spread. The stochastic processes are modelled by G2++ and CIR respectively.   
\item[3.]\textbf{Case 3 - Risk-free rates and default events}: we consider the option payoff in eq. \eqref{eqlook-backOpt}. The credit spread will be modelled via an inhomogeneous Poisson process, implying a determinist evolution of credit spread. This case considers default events (sampled from the credit spread implied default probability). In this case we can appreciate the impact of default on the liquidity spread estimation
\item[4.]\textbf{Case 4 - Risk-free rates, default events and credit spread}  Finally we are going to use the fully-fledged model, with stochastic default events and risk-free rates and credit spread evolution.
\end{itemize} 

\begin{figure}[!h]
    \centering
		\includegraphics[width=\textwidth]{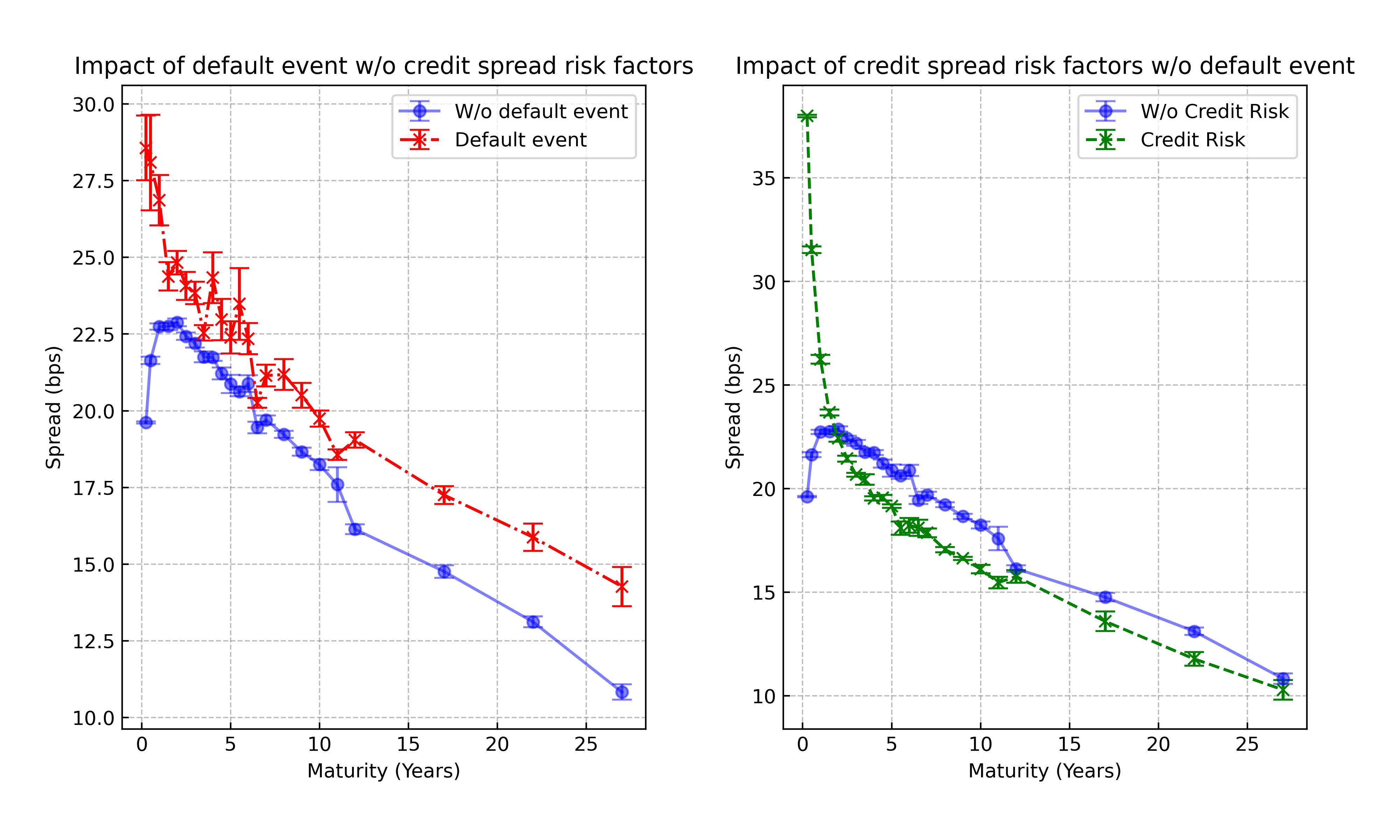}
    		\caption{Default events and credit spread volatility effects (BB rating). In blue is depicted the liquidity spread (expressed in bps) considering only the risk-free rates volatility effects (case 1). On the left in red is depicted the liquidity spread of case 2, while on the left in green is depicted the case 3.}
    		\label{fig:liqspreadcomparison}
\end{figure}

Let us compare case 1 with 2 and 1 with 3. This is depicted in Figure \ref{fig:liqspreadcomparison}. 

On the left of Figure \ref{fig:liqspreadcomparison}, we can see the impact of default events, case 1 vs case 3. Default events have a higher impact both on short-term ZC bonds and longer-term ZC bonds. In the first instance, short-term ZC bonds, the impact is due to the low number of probing days for the illiquid option. A default event excludes some probing dates from the maximization in the trajectory in which it occurs, from $\tau$ onwards. Given the fewer probing dates, this cut has a more significant impact than the case with more probing dates of a bond with a longer time to maturity. In the latter, long-term ZC bonds, given that we are considering a rated BB ZC bond, the number of default events is higher after a decade, resulting in a higher impact on liquidity spread estimation. In other words, the first effect (on short-term bonds) is due to the small number of probing dates considering  the short time to maturity; while the latter (on long-term bonds) is  linked to an effective cap on probing dates driven by the high default probability, effectively cutting the trajectory considered. We also observe that the error of the liquidity spread significantly increases when considering default events, w.r.t. the risk-free case. We also note that considering the default events prevents the liquidity spread curve from attaining a local maximum, excluding the extremes, compared to the spread in "risk-free" case that instead goes to zero for zero maturity and long maturity.

On the right of Figure  \ref{fig:liqspreadcomparison}, we can gauge the impact of credit spread volatility, comparing case 1 (blue) with case 2 (green). As one can see, the liquidity spread is lower after two years maturity when considering the credit spread volatility. The lower liquidity spread is primarily due to the interference of the uncertainty of credit spread and risk-free rates. This could be due to the asymmetric distribution generated by the CIR model. While there is a cap to increases in ZC bond prices (due to the floor at zero of credit spread), there is no floor to decrease in the ZC bond prices (higher credit spread). However, this is not the case for short-maturity ZC bonds. This is due to the behaviour of the credit spread in the first two years: the credit spread at time zero is well below its long-term value $\theta$. This implies a more significant and directional impact on the bond price, yielding a higher liquidity spread. This effect is present even if the credit spread at time zero is above the long-term value. Effectively, the credit spread evolution excludes the first probing dates since the bond value is affected by high credit spread. 

\begin{figure}[!h]
    \centering
		\includegraphics[width=0.8\textwidth]{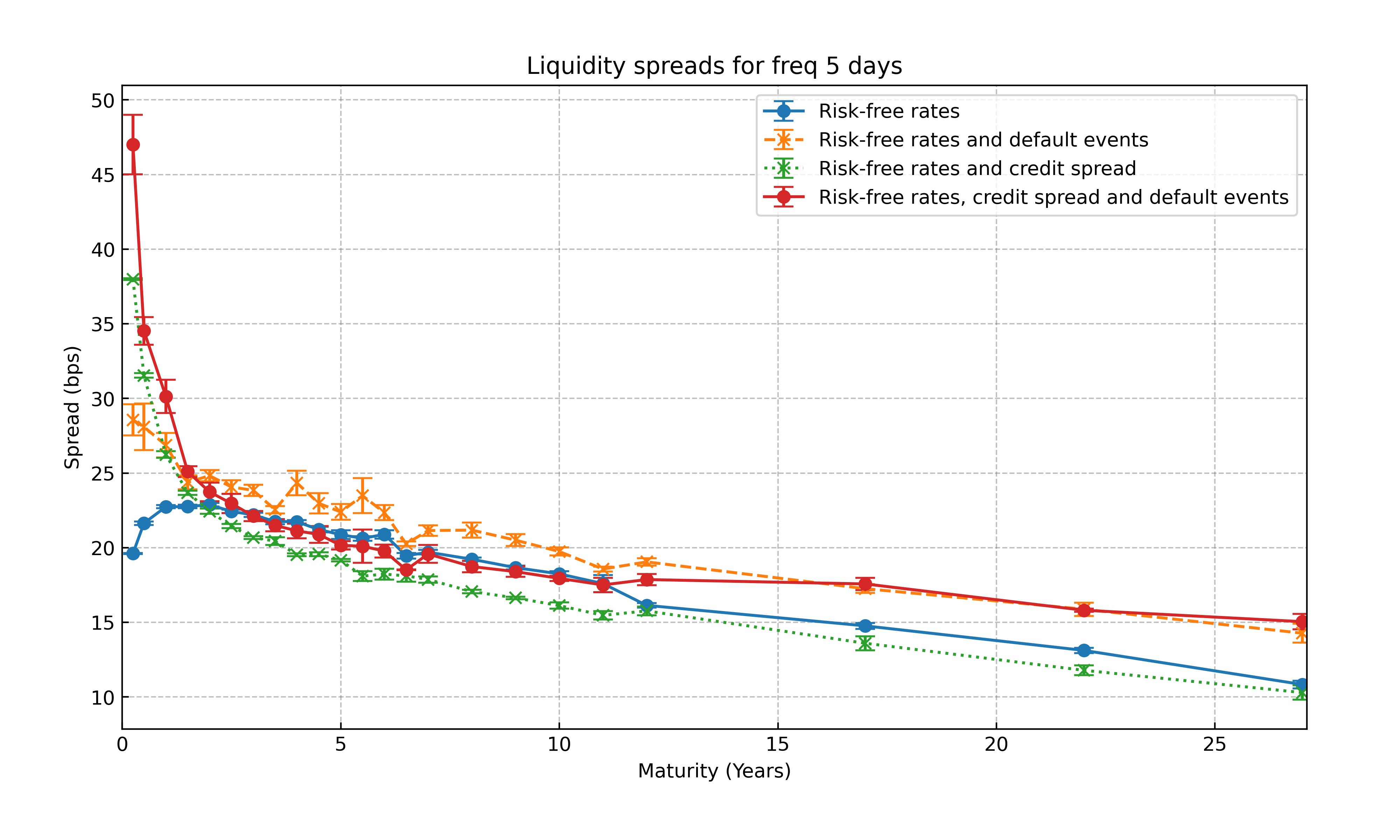}
    		\caption{Default events and credit spread volatility effects (BB rating). The four cases listed above are reported. }
    		\label{fig:liqspread10days}
\end{figure}

In Figure \ref{fig:liqspread10days}, we compare the four cases. We can observe that including the default events increases the liquidity spread estimation. However, this impact could be offset if interference between the credit spread volatility and the risk-free rate volatility exists. This can be seen when comparing the green dotted line (case 4) with the orange line (case 1). In the long term, the default event has a greater impact than everything else.

\begin{figure}[!h]
    \centering
		\includegraphics[width=0.8\textwidth]{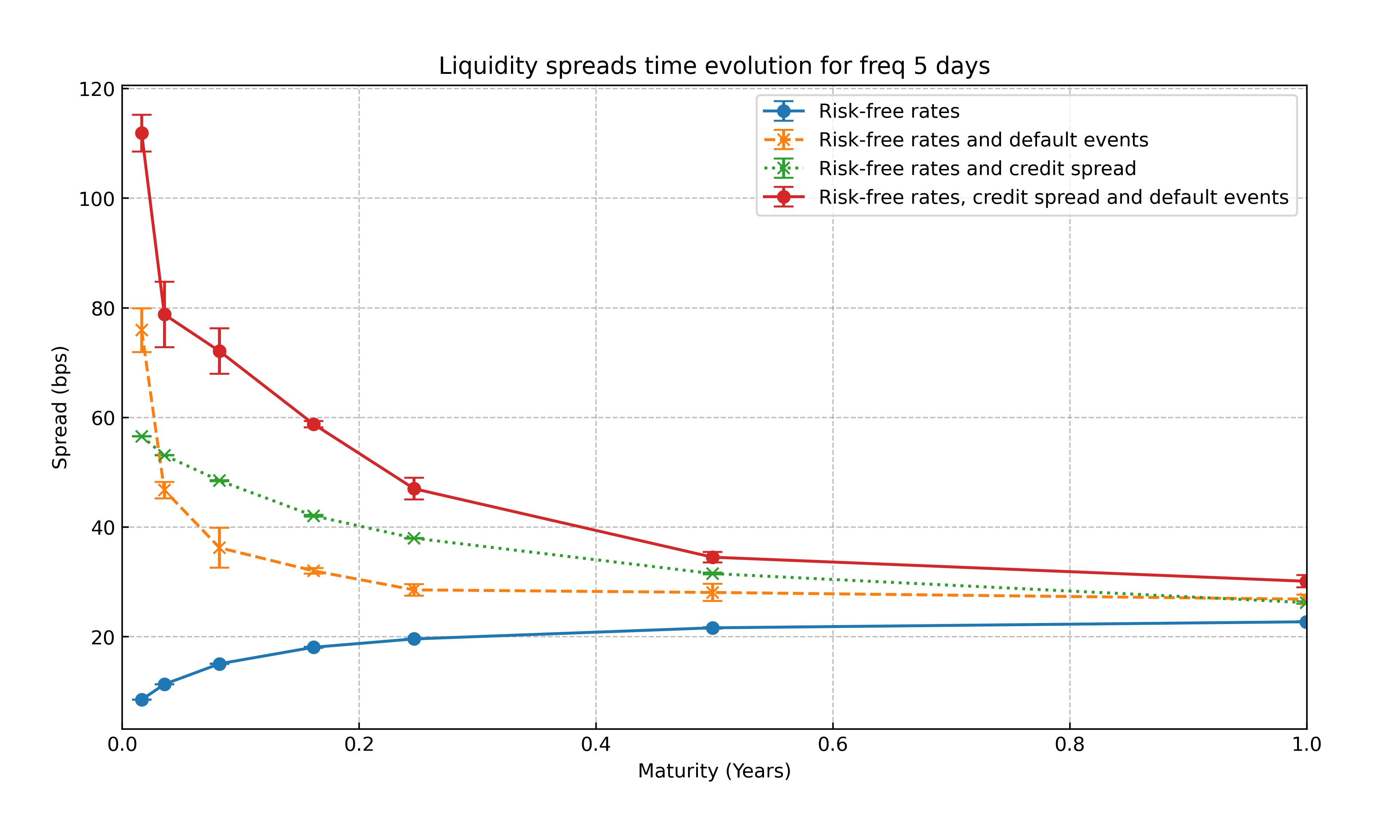}
    		\caption{Liquidity spread for short-maturity bonds  (BB rating)}
    		\label{fig:liqspreadshortmat}
\end{figure}

We can observe that, as in \cite{Koziol2003}, in general the liquidity spread decreases for longer maturity. The short maturity liquidity spread instead exhibit a new behaviour: the shorter the maturity the greater the impact of defaults. This is shown in Figure \ref{fig:liqspreadshortmat} where we can see that the liquidity spread curve attains its maximum near zero. This kind of behaviour is influenced by the level of default probability. In fact when considering a better rating (e.g. BBB) we can still observe the liquidity spread curve attaining a maximum for maturities greater than zero, as is displayed in \cite{Koziol2003} and it can be seen looking at the liquidity spread curve without considering credit risk and and default events. Notice that the increasing of credit spread for shorter maturity is not in contrast with the fact that liquidity price adjustments go to zero as the bond maturity goes to zero.  

\begin{figure}[!h]
    \centering
		\includegraphics[width=0.8\textwidth]{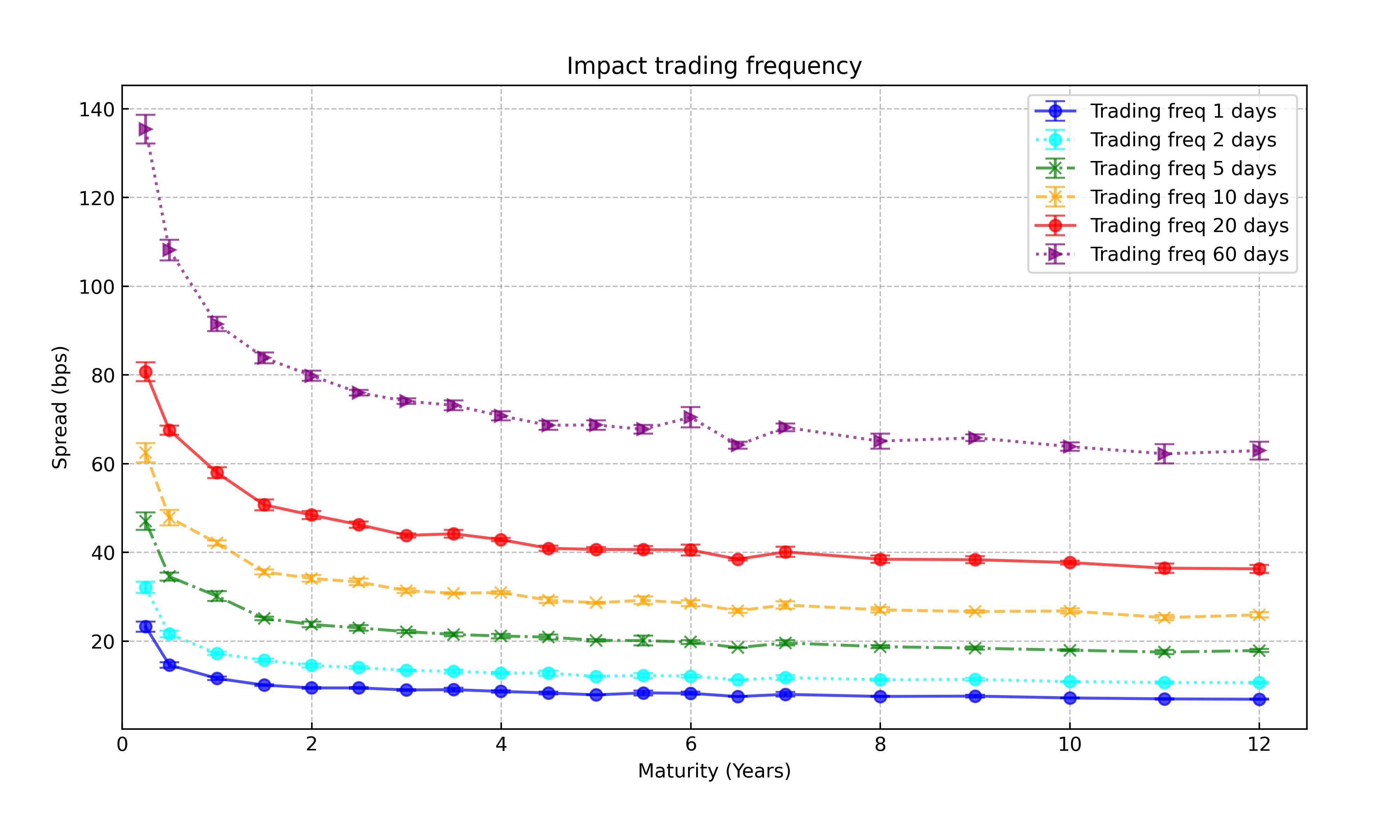}
    		\caption{Impact of probing frequency on the liquidity spread estimation (BB rating). The liquidity spread has been estimated in the case 4 }
    		\label{fig:liqtradfreq}
\end{figure}

In Figure \ref{fig:liqtradfreq} is depicted the impact of varying the probing frequency for the illiquid option. As expected the higher the trading frequency, the higher the liquidity spread.  

\section{A market application: estimating liquidity spread of an unquoted Republic of Italy bond}\label{sec4}

Let us give an example of how we can use the model to estimate a liquidity spread for an unquoted coupon-bearing bond.
Estimating liquidity spread for unquoted securities entails calibrating the probing frequency to estimate the illiquid option. In principle, one could rely on expert judgment to set this parameter. This is the case when there are not enough quoted bonds of the same issuer or comparables. When, instead, there are other quoted emissions, one can calibrate the probing frequency from market data. We are going to illustrate one procedure to estimate a liquidity spread for pricing a hypothetical emission of the Republic of Italy, whose features are summarized in Table \ref{table:bondItaly}.

\begin{table}[h]
\caption{Case bond feature}
\label{table:bondItaly}
\begin{center}
\begin{tabular}{lc}\toprule
\bf Contract feature&\bf Value\\ \midrule

Issue date&15/11/2023\\
Maturity date&15/11/2027\\
Ref. date&31/05/2024\\
Time to maturity&3.5 years\\
Coupon type&Fixed\\
Coupon Frequency& six month\\
Daycount convention and adjustment & $30/360$ and modified following\\
Coupon level&$4.5\%$\\
Amortizing type & Bullet\\
\bottomrule
\end{tabular}
\end{center}
\end{table} 

In order to estimate the value of the liquidity spread of the coupon bond, we need not only to calibrate to the market a model for the risk-free rates and the borrower credit spread (the Rep. of Italy), but we also need to specify the probing frequency of the illiquid option valuation (assuming that the liquid market trades every hour). Here, one can either set the probing frequency by expert judgment or try to estimate the probing frequency from the quoted emission. This estimation is, in turn, what we want to achieve.

To estimate the probing frequency yielding the liquidity spread of illiquid quoted bond, firstly, we will need a procedure to define which emissions can be considered liquid and which cannot. From this classification we then need to infer the liquidity spread for each illiquid bond and then calibrate the probing frequency of the model to correctly estimate the liquidity spread. For simplicity, this extra spread will be estimated as the difference between the yield curve calibrated on liquid bonds and the yield of the illiquid bond. The correct liquidity spread estimation should consider estimating the ZC liquidity spread that matches illiquid bond prices.

\begin{figure}[!h]
    \centering
		\includegraphics[width=0.8\textwidth]{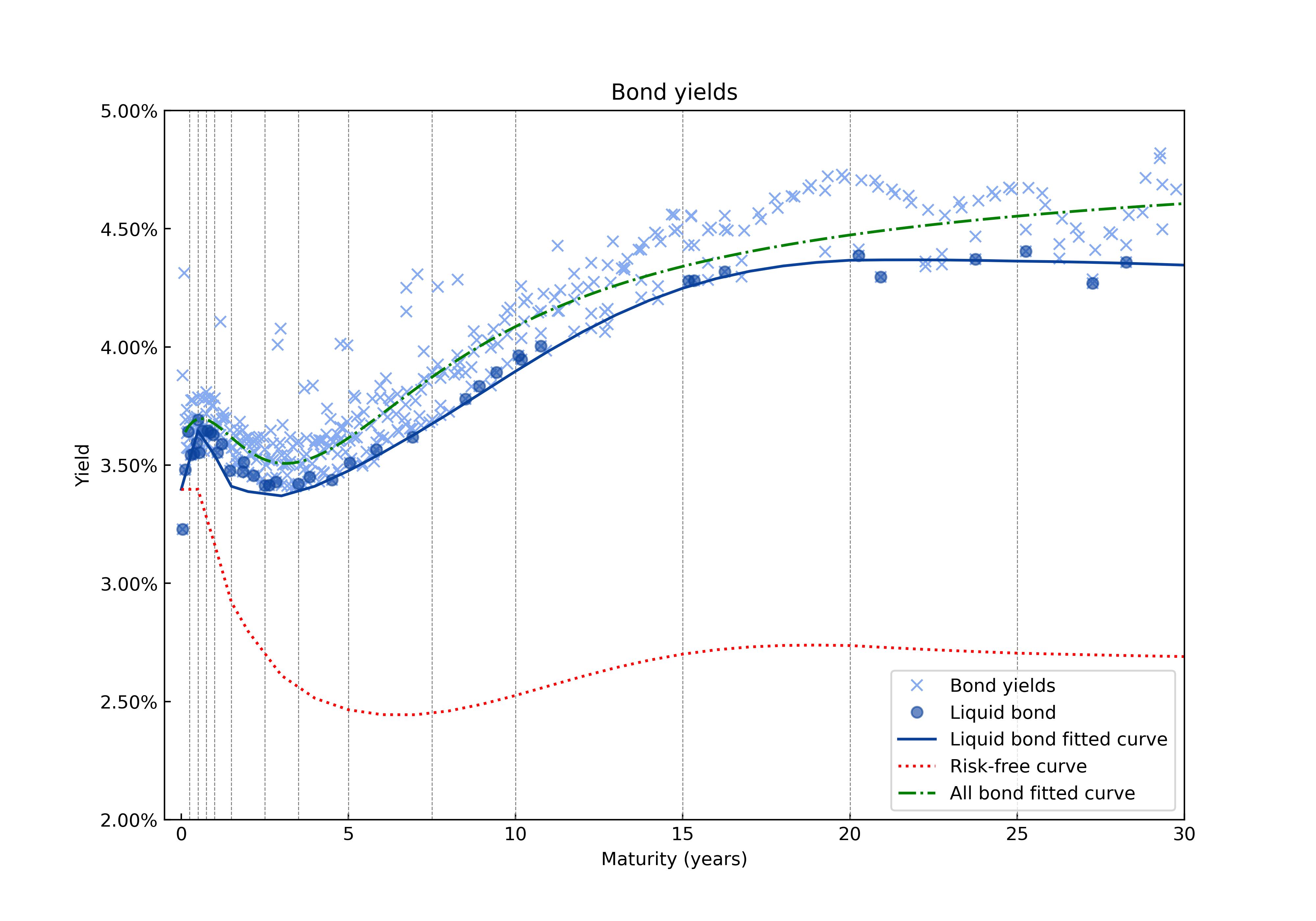}
    		\caption{Liquid bond selection and Liquid Yield curve fitting. Vertical dotted grey lines denote the time bucket edges.}
    		\label{fig:calibbond}
\end{figure}

We will rely on common liquidity indicators to classify quoted emissions: trading volume and bid-ask spread. The bid-ask spread depends on the time to maturity of the emission, since  duration mainly influences the sensitivity of the bond yields and prices. Therefore, we will define a bucket of time to maturity (see Table \ref{table:timebucket}) and select within each bucket the bond with a trading volume bigger than zero (in the last day). We will then rank the bonds from the smallest to the largest bid-ask spread. Ranking bonds in each bucket will allow us to select a few representatives as liquid bonds, those with the smallest bid-ask spread (we will choose three for each bucket). We will then build the liquid yield curve, fitting liquid market bond prices and then computing the yield curve. In picture \ref{fig:calibbond} in blue the selected liquid bond and the "Liquid" Yield curve is obtained. As expected the liquid yield curve is below the Yield curve obtained fitting the Svensson curve to  the whole sample \cite{Svensson1994}. Our selected bond under analysis has a time to maturity of 3.5 yeas and hence belongs to bucket G.

\begin{table}[h]
\caption{Time bucket definition}
\label{table:timebucket}
\begin{center}
\begin{tabular}{cc}\hline
\bf Bucket Name&\bf Time to Maturity (years)\\\hline
 A&0-0.25\\
B&0.25-0.5\\
C&0.5-0.75\\
D&0.75-1\\
E&1-1.5\\
F&1.5-2.5\\
G&2.5-3.5\\
H&3.5-5\\ 
I&5-7.5\\
L&7.5-10\\
M&10-15\\
N&15-20\\
O&20-25\\
P&25-30\\
Q&30-40\\
R&40$>$\\
\hline
\end{tabular}
\end{center}

\end{table}

\begin{figure}[!h]
    \centering
		\includegraphics[width=0.8\textwidth]{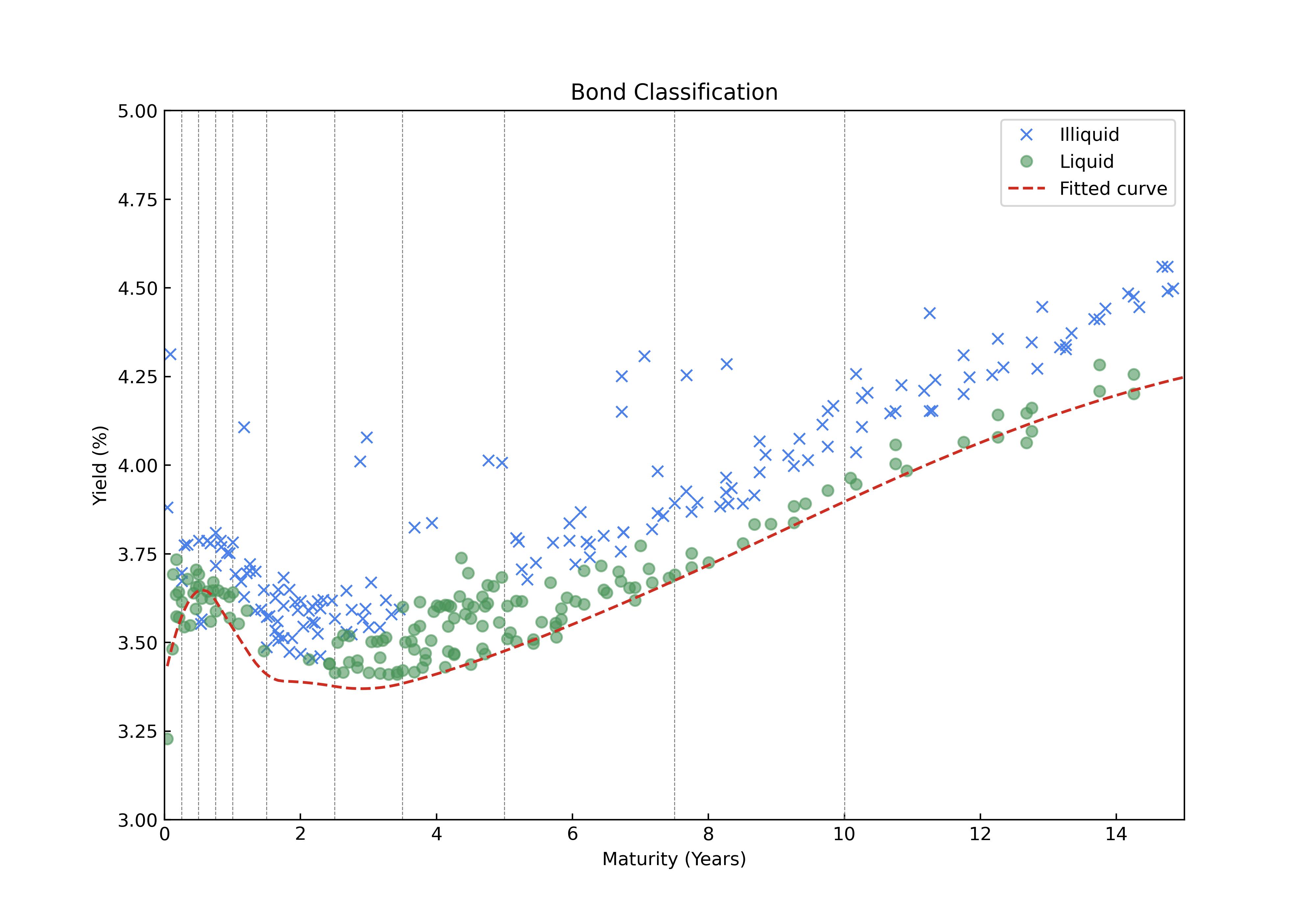}
    		\caption{ Liquid and illiquid bond calssification. Vertical lines denote the time bucket edges. }
    		\label{fig:bondclass}
\end{figure}

Given the "Liquid" yield curve, it is possible to identify the illiquid bond by computing the standard deviation of differences between the bond yields and the liquid yield curve for each bucket. We will classify illiquid bonds, those with more than one standard deviation from the Liquid yield curve, and their liquidity spread as their yield difference from the Liquid yield curve. We observe that this definition of illiquid bond is sensible to bid-ask spread or trading volume via the definition of the Liquid yield curve, but it also allows to assess the liquidity of bonds that have not been traded in the last day. Moreover, considering bonds with yield spread above one standard deviation allows to filter out the natural noise due to trading and selects the bond which are considered illiquid by the market. 
Notice also that the splitting of the yield curve in temporal bucket allows to gauge independently the liquidity of different sections of the yield curve. This can be seen in \ref{fig:bondclass}, where the time bucket F from 1.5 an 2.5 years exhibits a lower volatility than the adjacent bucket G and in turn has more illiquid bond. This difference in the number of illiquid bond could be due to calibration effects of the liquid yield curve. We can see that there is a trade-off when selecting temporal buckets: the finer the filtration on time to maturity, the more the ability to distinguish noise variation along the yield curve, but on the other hand the less homogeneous definition of liquid bond.   

We observe that this procedure could in principle yield for some bonds a negative liquidity spread (see e.g. bond in time bucket C). This happens when there has been a shift  in the yield curve of the issuer and an illiquid emission price did not yet adjust accordingly. In these cases, the model could help to predict which could be the current value of the bond, associated with the "new" yield curve and considering a liquidity spread comparable with other illiquid bond.\footnote{What we encounter can be consider a shortcoming of the procedure that is trying to detect the liquidity spread from quoted yield: it is in fact not easy to disentangle the misalignment of yield due to liquidity and those related to shift of the credit curve. To avoid this shortcoming one can filter the bond considering only those with trading volume in the last day (or week) different from zero, restricting the sample on which the analysis is performed.}

Once we have classified illiquid bonds, we can now calibrate the implicit probing frequency in the market quotes yielding the observed liquidity spread.

 The calibration of the probing frequency entails estimating the parameters for the model of risk-free rates and credit spread. In order to get sensible credit spread of the issuer (credit spread bigger than zero) we need to consider as risk free curve the yield curve of German BUND, because we are dealing with sovereign bonds. The BUND yield curve is depicted in red in Figure \ref{fig:calibbond}.  However this choice of risk-free curve poses a calibration problem for the G2++ model, since there are no quoted swaptions or cap and floor on bond yields. Nonetheless we calibrate on swaptions, assuming the volatility of swap rates is identical to the volatility of bond yields. The calibration of the CIR model parameters is then performed on the credit spread curve derived using the risk-free BUND curve. The calibrated parameters are reported in appendix A in Table \ref{table:g2ppmodelbund} and  \ref{table:CIRparameterIta}. The risk-free curve is reported in Table \ref{table:bundrisk-free}. 

The results for the first seven time buckets are reported in appendix B table \ref{table:calibLiqSpread}. 

\normalsize
For most cases, the calibration yields the probing frequency reproducing the market liquidity spread. Let us briefly comment on the few cases in which the calibration yields a suboptimal result. We can divide the cases in which the model fails to reproduce the market spread into two instances. In the first instance (happening for bonds \#8 and \#12 in time bucket C), the Market Liquidity spread is negative, and our model can reproduce only positive spreads. The prices and yield of these bonds are not aligned with current market conditions due to their illiquidity. The second instance, represented by bond \#1,\#2 (bucket A), and \#24 (bucket E), instead presents liquidity spread far greater than peer liquidity spreads. This can be due to misalignment of their price to current market conditions. However, in principle, our model should be able to reproduce their spreads; its failure can stem from the hypothesis of equal spacing between probing frequencies. As noted already in \cite{Koziol2003}, different dispersion of probing dates yields different liquidity spread estimations. We observe that this degree of freedom of the model, the choice of the probing dates, can be considered one of its limitations. 

In order to choose the probing frequency to estimate the liquidity spread of the analyzed bond reported in Table \ref{table:bondItaly}, we could extract different statistics from the probing frequency distribution for the illiquid bond in the same time bucket G where it belongs. In particular, we could choose a given percentile of the distribution. However, note that two outliers could be ruled out as illiquid bonds with unreliable liquidity spread estimation (they are more than 2 standard deviations from the mean of the whole sample and more than 40 standard deviations from the sample's mean, excluding these two outliers). Therefore, we are going to disregard bond \#64 and bond \#65. We will use the liquidity spread considering one standard deviation from the mean probing frequency, two standard deviations, and the maximum probing frequency. In other words, we will compute the liquidity spread for  14, 17, and 19 days. Considering different probing frequencies allows us to define a range of values for the liquidity spread. In case one must estimate the price for the bond, we believe that the two standard deviation estimate is a reasonable choice.

With these probing frequency we get that the liquidity spreads for our bond is as reported in Table \ref{table:liqEstimation}.
\begin{table}[!h]
\caption{Liquidity spread estimation of unquoted bond}
\label{table:liqEstimation}
\begin{center}
\begin{tabular}{ccc}
\toprule
\bf Distribution ref.&\bf Probing Frequency (days)&\bf Liquidity spread (bps)\\\midrule
1 std dev.&14&23\\
2 std dev.&17&24\\
Sample max.&19&27\\
\bottomrule
\end{tabular}
\end{center}
\end{table}

Let us conclude this section with a general comment. Market data are generally reliable when the market is liquid, and the quotes reflect the information available on the risk and return of securities. Instead, in illiquid market conditions, the reliability of the information extracted from quoted prices can not reflect the current market sentiment. Therefore, illiquid prices may not represent the correct prices. The abovementioned procedure aims to reduce errors due to a lack of updated information and identify potential outliers. Nonetheless, in some cases, it is not surprising that the quoted prices of illiquid instruments are misaligned with liquid quotes. Once identified, these quotes should be neglected.

\section{Conclusion}\label{sec5}
In this paper, we have shown a possible extension of the model proposed by \cite{Koziol2003} on the liquidity spread of bonds. This line of research stems from the seminal work of Longstaff \cite{Longstaff1995}, and it tries to estimate liquidity discount by measuring the impact on security valuation of different marketability characteristics. The liquidity discount is measured as the ratio of two look-back options with different probing dates. This paper extends the model in \cite{Koziol2003}, incorporating the impact of credit risk when estimating the look-back option. Credit risk has a double impact on the estimation of the option: on one hand, credit spread uncertainty affects the value of the bonds; on the other hand, default events truncate the set of possible probing dates for the look-back option payoff. The theoretical framework of the model has been described in section \ref{sec2}.

The two effects have been studied numerically in section \ref{sec3}, considering a specific model for credit spread evolution (CIR model) and risk-free rates (G2++). The analysis showed how the default event, especially for short maturity, has an impact on the liquidity spread magnitude. In particular, for low ratings (e.g., BB), the liquidity spread does not vanish as the time to maturity approaches zero. This behaviour differs from the case without credit risk or when the issuer's creditworthiness is excellent, where the liquidity spread goes to zero as the time to maturity goes to zero. Another important observation is that risk-free rate volatility and credit spread volatility could interfere destructively, lowering the overall volatility of the look-back option value, and in turn the liquidity spread. This is also due to the different distribution and effects that risk-free rates and credit spread have due to the model selection. Further development could investigate the role of the correlation between credit spread and risk-free rates, or it can consider different models with different correlation of the rate term structure.

In section \ref{sec4}, we investigate how one can calibrate the model (particularly the probing frequency for the illiquid look-back option) to market data. We showed how one could classify liquid and illiquid emissions of the same issuer in order to estimate the liquidity spread for quoted bonds. This classification relies on liquidity indicators, such as bid-ask spread and volume. The estimation of quoted liquidity spread allows the construction of a distribution of probing frequency for quoted illiquid bond spreads and, in turn, allows the choice of a proper probing frequency for the unquoted emission. Here, we encounter one of the major limitations common to all the models that deal with liquidity: by definition, the quotes referring to illiquid emission are the ones that adjust to new information with delay, i.e. they are "stale prices". Sometimes, therefore, quotes of illiquid bonds could not reflect the current market expectation of the credit risk or risk-free rates, rendering the classification challenging. Nonetheless, we were able to overcome this difficulty thanks to the high number of emissions of the chosen issuer (Rep. of Italy), which allowed us at least to rule out those with apparent inconsistency with the current market condition. However, disentangling the effect of liquidity and the delay to the mutated market conditions could be tricky for an issuer with few emissions.

As shown in this paper, our model could help study the effects of liquidity on bonds with different structures, coupons, amortizing schedules, etc., as well as the interaction between market conditions and liquidity. 
We conclude by underlying how this model can also be beneficial for unquoted emissions when a standard disposal period can be set by expert judgment. 
In this case the model could be calibrated on comparable bonds of the same industry with similar characteristics and the corresponding liquidity premium could be easily estimated applying the described approach and assuming a reasonable hypothesis on the length of the liquidation period.

\pagebreak
\printbibliography
\pagebreak

\appendix
\section{Models Parameters}
The model parameters for the analysis carried out in section \ref{sec3} are shown in Table \ref{table:CIRparameter}, \ref{table:g2ppmodel} \ref{table:risk-freecurve}.

\begin{table}[!h]
\caption{CIR model parameters for BBB and BB rating credit spread}
\label{table:CIRparameter}
\begin{center}
\begin{tabular}{lcc}\toprule
\bf Parameter&\bf Rating BBB&\bf Rating BB \\ \midrule

$\kappa$&$0.4455$&$0.7288$\\
$\theta$&$0.0141$&$0.0224$\\
$\sigma$&$0.0705$&$0.1689$\\
$r_0$&$0.0001$&$0.0054$\\
\bottomrule
\end{tabular}
\end{center}
\end{table} 

\begin{table}[!h]
\caption{These are the parameters of the G2++}
\label{table:g2ppmodel}
\begin{center}
\begin{tabular}{lc}\toprule
\bf Parameter&\bf Value\\ \midrule
$a$&$0.0693$\\
$\sigma$&$0.0116$\\
$b$&$0.0531$\\
$\eta$&$0.0057$\\
$\rho$&$0.1209$\\
\bottomrule
\end{tabular}
\end{center}
\end{table} 

\small
\begin{table}[!h]
\caption{Risk-free curve}
\label{table:risk-freecurve}
\begin{center}
\begin{tabular}{ c c  }
\toprule
	\bf Date & \bf Value \\ 
\midrule
	29/12/2023 & $1.0000$ \\ 
	11/1/2024 & $0.9986$ \\ 
	5/2/2024 & $0.9959$ \\
	4/4/2024 & $0.9896$ \\
	4/7/2024 & $0.9802$ \\ 
	6/1/2025 & $0.9652$\\ 
	5/1/2026 & $0.9452$ \\ 
	4/1/2027 & $0.9264$ \\
	4/1/2028 & $0.9065$ \\ 
	4/1/2029 & $0.8859$\\ 
	4/1/2034 & $0.7799$ \\ 
	4/1/2039 & $0.6812$ \\ 
	4/1/2044 & $0.6075$ \\ 
	4/1/2049 & $0.5519$ \\
	5/1/2054 & $0.5070$ \\ \bottomrule
\end{tabular}
\end{center}
\end{table}
\normalsize
The model parameters calibrated for the analysis in section \ref{sec4} are reported in Table \ref{table:g2ppmodelbund}, \ref{table:CIRparameterIta}, \ref{table:bundrisk-free}.

\begin{table}[!h]
\caption{These are the parameters of the G2++ calibrated on the 31/05/2024}
\label{table:g2ppmodelbund}
\begin{center}
\begin{tabular}{lc}\toprule
\bf Parameter&\bf Value\\ \midrule
$a$&$0.0195$\\
$\sigma$&$0.0062$\\
$b$&$0.0193$\\
$\eta$&$0.0062$\\
$\rho$&$0.0962$\\
\bottomrule
\end{tabular}
\end{center}
\end{table} 

\begin{table}[!h]
\caption{CIR model parameters for credit spread of Rep. of Italy}
\label{table:CIRparameterIta}
\begin{center}
\begin{tabular}{lc}\toprule
\bf Parameter& \bf Value\\ \midrule

$\kappa$&$0.0018$\\
$\theta$&$0.01$\\
$\sigma$&$0.005$\\
$r_0$&$0.0003$\\
\bottomrule
\end{tabular}
\end{center}
\end{table} 

\small
\begin{table}[!h]
\caption{Risk-free curve from BUND yields}
\label{table:bundrisk-free}
\begin{center}
\begin{tabular}{ c c  }
\toprule
	\bf Times (Years) & \bf Value \\ 
\midrule
0&1\\
0.08&0.997\\
0.25&0.991\\
0.5&0.982\\
1&0.965\\
2&0.94\\
3&0.917\\
4&0.897\\
5&0.874\\
6&0.856\\
7&0.835\\
8&0.814\\
9&0.791\\
10&0.768\\
15&0.651\\
20&0.566\\
25&0.495\\
30&0.438\\
\bottomrule
\end{tabular}
\end{center}
\end{table}

\section{Liquidity spread calibration}
Below the probing frequency reproducing the market liquidity spread.

\tiny
\begin{longtable}{cccccc}
\caption{Probing frequency calibration to illiquid bond liquidity spread.}
\label{table:calibLiqSpread}\\
\toprule
\multirow{2}{*}{\bf Bond \#}&  \multirow{2}{*}{ \bf Time Bucket}&\bf Time to maturity &\bf Market Liquidity &\bf Model Liquidity & \bf Probing \\
&& \bf (Years)& \bf Spread (bps)& \bf Spread (bps)& \bf Frequency (days) \\ \midrule 
1&A&0.04&45&8&8\\ 
2&A&0.09&84&12&16\\ \midrule
3&B&0.34&17&17&10\\ 
4&B&0.3&18&18&12\\
5&B&0.26&10&10&3\\ 
6&B&0.26&12&12&4\\ 
7&B&0.26&10&10&3\\ \midrule
8&C&0.54&-10&5&1\\ 
9&C&0.63&14&14&5\\ 
10&C&0.51&14&14&6\\ 
11&C&0.68&14&15&6\\ 
12&C&0.54&-8&6&1\\ \midrule
13&D&0.84&18&18&10\\ 
14&D&0.96&20&20&12\\ 
15&D&0.75&19&17&9\\ 
16&D&0.92&19&17&8\\ 
17&D&0.83&19&19&10\\ 
18&D&0.75&10&10&3\\ \midrule
19&E&1.04&16&16&7\\ 
20&E&1.21&22&22&13\\ 
21&E&1.34&26&26&17\\ 
22&E&1.13&17&17&7\\ 
23&E&1.46&23&23&16\\ 
24&E&1.17&62&40&107\\ 
25&E&1&24&24&15\\ 
26&E&1.26&26&26&18\\ 
27&E&1.26&24&24&15\\ 
28&E&1.17&14&14&5\\ 
29&E&1.42&17&17&8\\ 
30&E&1.33&15&15&6\\ \midrule
31&F&1.96&20&20&12\\ 
32&F&2.46&24&24&14\\ 
33&F&1.96&20&20&12\\ 
34&F&2.46&24&24&16\\ 
35&F&2.04&16&16&7\\ 
36&F&1.54&17&17&7\\ 
37&F&2.21&17&17&8\\ 
38&F&1.84&26&26&16\\ 
39&F&2.34&24&24&16\\
40&F&1.63&23&23&17\\ 
41&F&2.12&20&20&10\\ 
42&F&1.68&13&13&4\\ 
43&F&2.17&22&22&11\\ 
44&F&2&23&23&14\\ 
45&F&1.51&16&16&7\\ 
46&F&2.26&23&23&13\\ 
47&F&1.75&29&29&27\\ 
48&F&1.75&21&21&11\\ 
49&F&2.29&21&21&11\\ 
50&F&2.26&14&14&5\\
51&F&2.17&17&17&7\\ 
52&F&1.68&26&26&14\\ 
53&F&1.92&22&22&13\\ 
54&F&1.66&17&17&6\\
55&F&2.29&8&7&2\\ 
56&F&1.88&12&12&5\\ 
57&F&1.63&14&14&5\\ 
58&F&2.17&7&7&2\\ 
59&F&1.84&8&8&2\\
60&F&1.68&11&11&3\\
61&F&2&8&8&2\\ 
62&F&1.51&8&8&2\\ 
63&F&1.75&12&12&4\\ \midrule
64&G&2.88&64&64&175\\ 
65&G&2.98&71&59&217\\ 
66&G&2.96&23&22&10\\ 
67&G&3.46&21&21&11\\ 
68&G&3.34&20&20&9\\ 
69&G&2.67&16&16&6\\ 
70&G&3&17&17&8\\ 
71&G&2.51&19&19&9\\ 
72&G&2.75&22&23&12\\
73&G&3.26&24&24&14\\ 
74&G&2.75&15&15&6\\ 
75&G&2.67&27&27&19\\ 
76&G&3.17&17&17&7\\ 
77&G&2.92&20&20&10\\ \bottomrule

\end{longtable} 

\end{document}